\documentclass{article}
\usepackage{arxiv}
\usepackage[utf8]{inputenc} 
\usepackage[T1]{fontenc}    
\usepackage{hyperref}       
\usepackage{url}            
\usepackage{booktabs}       
\usepackage{amsfonts}       
\usepackage{nicefrac}       
\usepackage{microtype}      
\usepackage{lipsum}
\usepackage{graphicx}
\usepackage{caption}
\usepackage{tabto}
\usepackage{multirow}
\usepackage{makecell}
\usepackage{amsmath}
\usepackage{titlesec}
\usepackage{float}
\usepackage{cite}
\usepackage{cleveref}
\usepackage{array}
\newcolumntype{P}[1]{>{\centering\arraybackslash}p{#1}}

\usepackage[title]{appendix}
\graphicspath{ {./images/} }

\title{Enhancing Disruption Prediction through Bayesian Neural Network in KSTAR}

\author{
 Jinsu Kim \\
  Department of Nuclear Engineering\\
  Seoul National University\\
  \texttt{wlstn5376@gmail.com} \\
   \And
 Jeongwon Lee \\
  Korean Institute of Fusion Energy\\
  \texttt{jeongwonlee@kfe.re.kr} \\
  \And
 Jaemin Seo \\
  Department of Physics\\
  Chung-Ang University \\
  \texttt{jseo@cau.ac.kr} \\
  \And
Young-Chul Ghim \\
  Department of Nuclear and Quantum Engineering\\
  Korea Advanced Institute of Science and Technology \\
  \texttt{ycghim@kaist.ac.kr} \\
  \And
 Yeongsun Lee \\
  Department of Nuclear Engineering\\
  Seoul National University\\
  \texttt{leeys1996@snu.ac.kr } \\
  \And
Yong-Su Na \\
  Department of Nuclear Engineering\\
  Seoul National University\\
  \texttt{ysna@snu.ac.kr} \\
}

\begin{document}
\NumTabs{16}

\maketitle
\begin{abstract}
\tab Disruption in tokamak plasmas, stemming from various instabilities, poses a critical challenge, resulting in detrimental effects on the associated devices. Consequently, the proactive prediction of disruptions to maintain stability emerges as a paramount concern for future fusion reactors. While data-driven methodologies have exhibited notable success in disruption prediction, conventional neural networks within a frequentist approach cannot adequately quantify the uncertainty associated with their predictions, leading to overconfidence. To address this limit, we utilize Bayesian deep probabilistic learning to encompass uncertainty and mitigate false alarms, thereby enhancing the precision of disruption prediction. Leveraging 0D plasma parameters from EFIT and diagnostic data, a Temporal Convolutional Network adept at handling multi-time scale data was utilized. The proposed framework demonstrates proficiency in predicting disruptions, substantiating its effectiveness through successful applications to KSTAR experimental data.
\end{abstract}

\section{Introduction}
\tab One of the prominent challenges in fusion research is to avoid and suppress disruption, which is an abrupt and uncontrolled termination process of plasma resulting from various plasma instabilities and control errors \cite{zakharov2012understanding, schuller1995disruptions, wesson1989disruptions, de2011survey}. During this process, a large amount of energy and the loss of particles are released, and this leads to detrimental effects on tokamak devices, including severe structural damage to the walls and divertors. This is undesirable on a reactor-relevant scale like ITER since the existing materials cannot endure the deconfined heat. Thus, predicting disruptions in advance is crucial for establishing appropriate strategies to avoid and mitigate disruptions. However, there are still challenges not only in understanding its complex dynamics \cite{zhu2020hybrid} but also in designing for real-world implementation. For instance, a real-time disruption predictor must identify disruptions prior to the minimum required prediction time of 40 ms before thermal quench \cite{hollmann2015status}, but the prediction error can be substantial in such cases, potentially misleading the prediction system as the prediction time increases. Therefore, the uncertainty of disruption prediction should be considered to ensure better confidence in real-world implementation.

\tab There are two distinguished ways to predict disruptions in fusion research: Physics-based approach \cite{9004540, sabbagh2018disruption, sabbagh2023disruption} and data-driven approach \cite{cannas2007prediction, kates2019predicting, vega2015disruption, ferreira2020deep, churchill2020deep, zheng2023disruption, rea2018disruption, zhu2020hybrid, vega2022disruption, chandrasekar2020data, guo2021disruption}. The former has good interpretability and physical consistency for predicting disruptions through physical models combined with MHD theory. The latter includes deep learning, a subset of data-driven methods rooted in neural computing, and has gained prominence due to its remarkable performance. Guaranteed by the universal approximation theorem \cite{hornik1989multilayer}, neural networks are recognized as effective approximators for nonlinear mapping functions in various tasks, leading to the potential to address the challenges associated with disruption prediction in fusion research.

\tab Despite its potential, deep learning has two critical issues: Overfitting \cite{ying2019overview} and Overconfidence \cite{hein2019relu, wang2021rethinking}. Overfitting is related to low generalization due to the tendency of fitting closer to the training data than to the underlying distribution. These can be mitigated through data augmentation or reducing the model complexity. Meanwhile, Overconfidence is when a model exhibits excessive certainty or confidence in its output, even when the output is incorrect or uncertain. Several factors, including over-parameterization, insufficient data, model complexity, and so on, result in overconfidence, and this misleads the disruption prediction system due to excessive confidence in its incorrect predictions, including a false alarm or a missing alarm. To cover this, Bayesian deep learning \cite{blundell2015weight} is utilized. This provides a principle framework for modeling uncertainty by applying a stochastic neural network, allowing the integration of prior knowledge and facilitating a more robust estimation of model parameters. By modeling uncertainty, neural networks can gain the capacity to quantify the uncertainty in their predictions, thereby mitigating overconfidence and enhancing generalization to unseen data. Furthermore, utilizing uncertainty contributes to the overall reliability of its prediction and interpretability, consequently enhancing the safety and reliability of the disruption prediction system.

\begin{figure}[t]
    \centerline{\includegraphics[width=15cm, height = 10cm]{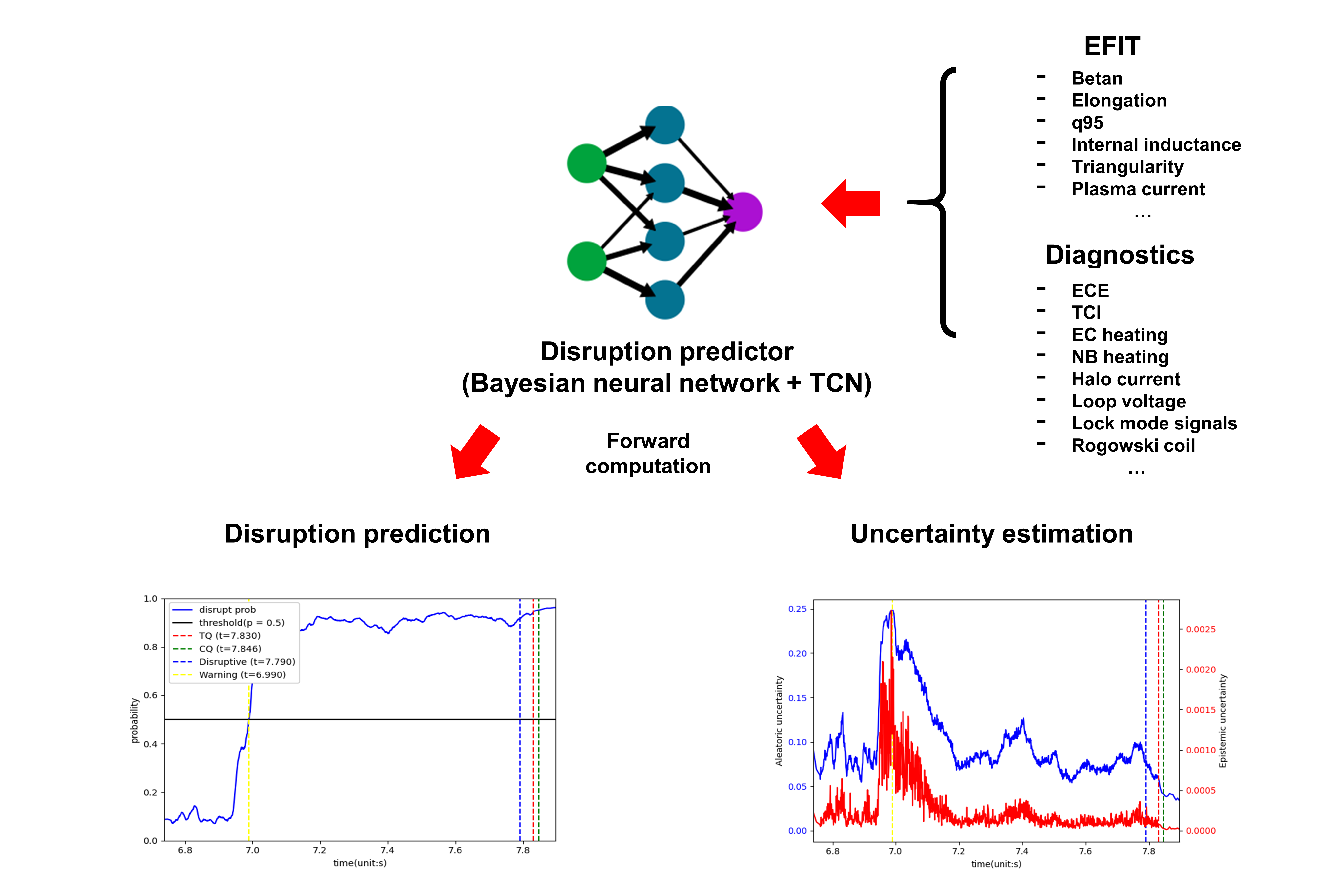}}
    \caption{The proposed framework of our work. The TCN architecture allows the model to handle long temporal sequence data while the Bayesian neural network predicts disruption with uncertainty estimation in the inference process.}
    \label{fig:framework}
\end{figure}

\tab In this research, we introduce an integrated predictor based on Bayesian neural networks designed to facilitate the disruption probability and the associated uncertainty, as described in Figure \ref{fig:framework}. Our objective is to achieve an accurate disruption prediction prior to the minimum required prediction time of 40ms before thermal quench \cite{hollmann2015status}. We leveraged 0D plasma parameters derived from EFIT and diagnostics in conjunction with a Temporal Convolutional Network (TCN) architecture \cite{oord2016wavenet}, which can efficiently handle multi-scale long sequential data compared with recurrent neural networks such as LSTM and GRU. This was demonstrated to be adequate for disruption prediction with Electron Cyclotron Emission imaging diagnostic (ECEi) \cite{churchill2020deep}. The TCN architecture in our model encodes multi-scale temporal input signals from EFIT and diagnostics. Next, a Bayesian neural network classifies the plasma state with encoded input signals from the TCN architecture and computes the uncertainty associated with its output probability. This uncertainty quantification proves instrumental in enhancing the accuracy of disruption prediction by enabling the adjustment of thresholds for the model's output and its associated uncertainty. By synthesizing the abovementioned components, our integrated disruption predictor presents a comprehensive and advanced approach to predicting disruptions. 

\tab The remainder of this paper is structured as follows. The dataset construction is described in section 2. In section 3, the proposed methodologies that we used are presented. Then, the evaluation of the model and the analysis to demonstrate the validity of our proposed method is covered in section 4. Finally, our research is concluded in section 5.

\section{Dataset Construction}

\tab The physical mechanisms leading to disruptions have been investigated by several studies \cite{zakharov2012understanding, schuller1995disruptions, wesson1989disruptions, de2011survey}, yet a comprehensive understanding remains challenging. Nonetheless, it is essential to recognize the root causes of disruptions that induce instabilities in tokamak plasma to achieve disruption prediction. Wesson et al. \cite{wesson1989disruptions} have outlined that disruptions at JET can be mainly categorized as (1) Density limit disruption, (2) Low q-limit disruption, (3) Current rise disruption, and (4) Vertical instability disruption. These classifications are rooted in the causes or mechanisms that give rise to disruptions, providing a structured framework for understanding the diverse nature of disruptions in other devices. This study considered these major disruptions and their mechanisms to select disruption-relevant signals to identify disruption precursors in the KSTAR tokamak device. Detecting operational limits, such as density limit and low-q disruptions, requires radiative information, electron temperature and density, and q profile. Electron cyclotron emission (ECE) and magnetic diagnostic signals are also essential to offer trigger signals for identifying disruptions \cite{wesson1989disruptions}. This choice is substantiated by the observation that the minor disruption, leading to major disruption, involves a fast radial redistribution of the plasma energy that can be detected by diagnosis of electron temperature profiles through ECE measurement. Furthermore, magnetic diagnostic signals can identify the magnetic perturbation induced by the magnetic island \cite{wesson2011tokamaks}. 

\begin{table}[H]
\captionof{table}{List of signals used for the construction of dataset}
\begin{center}
\begin{tabular}{p{3cm}|c}
\hline
\multicolumn{1}{c|}{\textbf{Channel}} & \textbf{Description}      \\ \hline
\multicolumn{1}{c|}{}            & Plasma current                 \\
\multicolumn{1}{c|}{}            & Normalized beta                \\
\multicolumn{1}{c|}{}            & Poloidal beta                  \\
\multicolumn{1}{c|}{}            & Elongation                     \\
\multicolumn{1}{c|}{}            & Safety factor                  \\
\multicolumn{1}{c|}{EFIT}        & Major radius                   \\
\multicolumn{1}{c|}{}            & minor radius                   \\
\multicolumn{1}{c|}{}            & Internal inductance            \\
\multicolumn{1}{c|}{}            & Triangularity                  \\
\multicolumn{1}{c|}{}            & Toroidal magnetic field        \\
\multicolumn{1}{c|}{}            & R position of X-point          \\
\multicolumn{1}{c|}{}            & Z position of X-point          \\ \hline
\multicolumn{1}{c|}{}            & Electron cyclotron emission    \\
\multicolumn{1}{c|}{}            & Lock mode signal               \\
\multicolumn{1}{c|}{}            & Halo current monitoring signal \\
\multicolumn{1}{c|}{}            & Diamagnetic loop               \\
\multicolumn{1}{c|}{}            & TCI:line-averaged electron density \\
\multicolumn{1}{c|}{Diagnostics} & Loop voltage                   \\
\multicolumn{1}{c|}{}            & H alpha signal                 \\
\multicolumn{1}{c|}{}            & EC heating power               \\
\multicolumn{1}{c|}{}            & NB heating power               \\
\multicolumn{1}{c|}{}            & Rogowski coil                  \\
\multicolumn{1}{c|}{}            & Magnetic field probe           \\
\multicolumn{1}{c|}{}            & UV photodiode bolometer        \\ \hline
\end{tabular}
\end{center}
\label{table:dataset}
\end{table}

\tab Numerous studies utilizing machine learning or deep learning methodologies have incorporated EFIT data containing equilibrium information of plasma and magnetic diagnostic data such as Mirnov coil current and lock model signals \cite{vega2015disruption, zhu2020hybrid}. As highlighted, disruption prediction proceeding with the onset of thermal quench demands the consideration of diverse precursors derived from multiple input signals. Consequently, this research harnessed EFIT and various diagnostic signals, including disruption-relevant physical information, as described in Table \ref{table:dataset}.

\tab It is crucial to recognize discrepancies in the sampling rates among these signals for dataset construction. The time interval between data points for EFIT is 10 ms, while the interval is 1 ms for diagnostic data. The requisite number of data points for sequential input data of EFIT and diagnostics should be different to make each signal's temporal length uniform. The resulting dataset thus exhibits multi-modality due to the presence of multi-scale input signals, necessitating the accommodation of this complexity for training our disruption predictor. To address this issue, our proposed model contains multiple modules of TCN encoders that allow parallelized processing for different input data modalities. 

\tab The data labeling process for identifying the disruptive phase in KSTAR proceeded with the following details. The $I_p$ minimum fault time, where the plasma current fell below 200 kA, was utilized as the criterion for labeling the current quench time. The thermal quench time was manually annotated by comparing the timing of the detected peak value of the DMF signal and the point at which the ECE signals descended below 10 eV. The time corresponding to the thermal quench was denoted as $t_{disrupt}$, differentiating disruptive and non-disruptive data.

\section{Methodologies}

\tab Our research aims to predict the disruption in advance to thermal quench with uncertainty estimation for a reliable disruption alert. We designed the overall framework for disruption prediction to achieve our purpose, as shown in Figure 1. The suggested model adeptly handles multi-modal and multi-scale input data and computes the probability of disruption onset, concurrently estimating the uncertainty of its prediction. This capacity is significant because overconfidence, one of the main challenges of deep learning that harms the reliability of neural networks, can be addressed by estimating the uncertainty of their predictions. We will explain the details of the components used in this research in the following sub-sections.

\textbf{Temporal Convolution Network} A Temporal Convolutional Network (TCN) \cite{oord2016wavenet} is a type of neural network architecture designed for processing sequential data, particularly in the time domain. It utilizes one-dimensional convolutions to efficiently capture dependencies and patterns in sequential data through dilated causal convolutions. In more detail, causal convolution operations inhibit the model from violating the ordering of sequences during prediction. In contrast, dilated convolution operations use larger receptive fields without drastically increasing the number of parameters by adding gaps between filter elements. These allow the computationally efficient process for covering time series data and capturing long-range dependencies without a complex architecture. R.M. Churchill et al. \cite{churchill2020deep} have shown significant results with ECEi for predicting disruptions through TCN. In this work, TCN architecture is applied as a temporal encoder for our integrated model, which extracts the embedded hidden vectors from temporal input data to identify disruptions at the classifier layer. A noise layer that adds Gaussian noise to the input data to avoid overfitting is combined with TCN. Moreover, several parallel TCN architectures with different configurations are employed to handle multiple modalities of temporal data. The details of our integrated model are described in Appendix A.

\textbf{Bayesian Neural Network} A Bayesian neural network \cite{blundell2015weight} is a stochastic neural network that incorporates Bayesian probabilistic learning into its architecture. The main difference between a Bayesian neural network and other traditional neural networks is that model weights are random variables sampled from probability distributions. In contrast, those of conventional cases are deterministic values. This stochastic property allows the network to quantify uncertainty from model parameters and predictions. A Bayesian neural network's foundation lies in applying variational inference \cite{blei2017variational} to the neural network’s architecture. Unlike traditional cases, variational inference computes the posterior distribution of weights, which can be optimized by variational learning. The objective function for a Bayesian framework, the variational free energy, is expressed in Equation \ref{eq:ELBO}. 

\begin{equation}
F(D,\theta)=KL[q(w|\theta)|p(w)]-E_{q(w|\theta)}[\log p(D|w)]
\label{eq:ELBO}
\end{equation}

Unfortunately, it is computationally challenging to minimize this objective function directly. Thus, Charles Blundell et al. \cite{blundell2015weight} simplified the objective function by the re-parameterization and Monte Carlo approximation, also known as Bayes by Backprop. Equation \ref{eq:ELBO-approximation} expresses the approximated formula.

\begin{equation}
F(D,\theta) \approx \sum_{i=1}^{n}\log q(w^{(i)},\theta) - \log p(w^{(i)}) - \log p(D|w^{(i)})
\label{eq:ELBO-approximation}
\end{equation}

For each weight, the Gaussian variational posterior is utilized for sampling the weight $w$, which is parameterized by the mean and standard deviation. The classifier module of our integrated model contains a Bayesian linear layer, which can compute the probability of disruptions. We applied transfer learning for efficient learning. A disruption predictor with a non-Bayesian classifier is trained first. Then, the predictor with a Bayesian classifier is trained by transferring the weights of the encoder from the non-Bayesian disruption predictor. The required time for training a Bayesian disruption predictor can be reduced through this process. The advantages of a Bayesian neural network are multifaceted, including uncertainty quantification \cite{kendall2017uncertainties}, robustness to overfitting \cite{blundell2015weight}, and practical utility to deep learning practitioners \cite{gal2016dropout}. More reliable disruption prediction can be achieved using these advantages, which will be shown later.

\textbf{Uncertainty estimation} For a Bayesian neural network with input data $x$ and output data $y$, the predictive distribution $p_D(y|x)$ is given by $p_D(y|x)=\int{p_w(y|x)p_D(w)dw}$, where $w$ is an weight parameter of the model. The probability distribution of weight is estimated by Gaussian distribution $q_{\theta}(w|D) \sim N(w|\mu,\sigma^2)$ and it is possible to construct an estimation of expectation of predictive distribution by Monte Carlo approximation, meaning that sampling from $q_{\theta}(w|D)$ can approximate the integral formula of predictive distribution, as expressed in Equation \ref{eq:predicive-uncertainty}.

\begin{equation}
E_q[p_D(y|x)] = \int{p_w(y|x)p_D(w)dw} \approx \frac{1}{T}\sum_{j=1}^{T} p_{w_j}(y|x)
\label{eq:predicive-uncertainty}
\end{equation}

$T$ is defined as the number of samples. Using this approximation, evaluating the uncertainty of the model's prediction is now available, referred to as predictive variance. The predictive variance can be divided by two distinctive terms: Aleatoric uncertainty and Epistemic uncertainty. See Equation \ref{eq:aleatoric-epistemic-uncertainty}.

\begin{equation}
\begin{split}
    Var_q[p_D(y|x)] = E_q[yy^T] - E_q[y]E_q[y]^T=\underbrace{\int{[diag(E_p[y]) - E_p[y]E_p[y]^T]q_\theta (w|D)dw}}_{\text{Aleatoric uncertainty}} \\
    +\underbrace{\int{(E_p[y]-E_q[y])(E_p[y]-E_q[y])^Tq_\theta (w|D)dw}}_{\text{Epistempic uncertainty}}
\end{split}
\label{eq:aleatoric-epistemic-uncertainty}
\end{equation}

Here, aleatoric uncertainty is stochastic uncertainty induced by noisy data or the random nature of the system. Meanwhile, epistemic uncertainty is systematic uncertainty caused by the model or the lack of knowledge of the system. In the Bayesian approach, the variability of model weight $w$ results in epistemic uncertainty, and this quantity can be reduced by increasing the sample size. If data quality is low, there might be high aleatoric uncertainty, while the model has low capability when high epistemic uncertainty is observed. Similar to equation (3), these uncertainties can be estimated by Monte Carlo approximation. Yongchan Kwon et al. \cite{kwon2020uncertainty} proposed an innovative way to compute these uncertainties in multi-class classification described as Equation \ref{eq:aleatoric-epistemic-uncertainty-approximation}.

\begin{equation}
    Var_q[p_D(y|x)] = \underbrace{\frac{1}{T}\sum_{j=1}^{T} diag(\hat{p}_j)-\hat{p}_j \hat{p}_j^T}_{\text{Aleatoric uncertainty}} +\underbrace{\frac{1}{T}\sum_{j=1}^{T} (\hat{p}_j-\Bar{p})(\hat{p}_j-\Bar{p})^T}_{\text{Epistempic uncertainty}}
\label{eq:aleatoric-epistemic-uncertainty-approximation}
\end{equation}

where $p_j=p(w_j)$ is the output of the Bayesian model and $\Bar{p}$ is a mean value of the output. Our work utilized the proposed approximation to estimate these uncertainties effectively. Our analysis and optimizing process for enhancing the model's accuracy are based on these quantities obtained from the network.

\textbf{Training and Evaluation} A disruption predictor is required to discern whether the input data representing the plasma state aligns with a disruptive or non-disruptive state. Thus, it is apt to frame the objective of this work as a binary classification task. It is conventionally understood that setting the objective function or loss function for training a binary classifier involves using cross-entropy loss, as described in Equation \ref{eq:cross-entropy}.

\begin{equation}
L_{CE} = -\frac{1}{N}\sum_{t=1}^{N} {w_t [y_t \log({\hat{y_t}}) + (1-y_t) \log({1-\hat{y_t}})]}
\label{eq:cross-entropy}
\end{equation}

In equation (8), $y_t \in{\{0,1\}}$ represents the label, where the label 1 signifies the disruptive class, and the label 0 corresponds to the non-disruptive class. $\hat{y_t}$ denotes the output probability from the neural network. $w_t$ is a weighting factor representing the ratio between disruptive and non-disruptive data. However, cross-entropy loss has difficulty adequately addressing the challenge of class imbalance when many easy examples and few hard samples coexist \cite{lin2017focal}. This work utilized Focal Loss \cite{lin2017focal}, a modified cross-entropy loss designed to address the classification challenges posed by a sparse set of hard samples, as described in Equation \ref{eq:Focal-Loss}.  

\begin{equation} 
L_{Focal} = -\frac{1}{N}\sum_{t=1}^{N}{w_t [y_t (1-\hat{y_t})^\gamma \log({\hat{y_t}})+(1-y_t) \hat{y_t}^\gamma \log({1-\hat{y_t}})  ]}
\label{eq:Focal-Loss}
\end{equation}

Additionally, our model is equipped with a noise layer, which adds Gaussian random noise with a mean of 0 and a standard deviation of 1.0 to its input data as a precaution against overfitting. Throughout the training process, optimization is performed using AdamW in conjunction with a StepLR scheduler. To evaluate the model capability, True Positive Rate (TPR, a ratio between true alarm cases and all positive cases), False Positive Rate (FPR, a ratio between false alarm cases and all negative cases), F1 score (a harmonic mean of the precision and recall), precision (the ratio of true positives to predicted positives), recall (the ratio of true positives to all positives, ROC (Receiver Operating Characteristic) curve, and AUC (Area under the ROC Curve) are utilized. 

\textbf{Setting} The temporal sequence length of the input data is consistently set to 1000 ms for all cases. The EFIT data exhibit a time interval of 10ms, while diagnostic data maintain a time interval of 1ms. Consequently, the EFIT data comprises 100 data points, and diagnostic data includes 1000 data points for each input feature. The dataset encompasses experimental data from KSTAR campaigns conducted between 2019 and 2022, containing 301 shots in total. The dataset is divided into training, validation, and test sets by a random split, with respective ratios of  64\%, 16\%, and 20\%. The training epoch is 128 for all cases. All input data are transformed as multi-variable temporal sequential data, denoted by a data shape of $(B, T, D)$, where $B$, $T$, and $D$ represent the batch size, temporal length, and the number of input features, respectively. Considering the effect of the scales across the input features in model performance, Robust Scaler is utilized to reduce the sensitivity to input values and remove outliers.

\section{Results and Discussion}
\tab In this section, we evaluate our proposed model with quantitative metrics to check the model performance. To analyze the model capabilities, we utilize t-SNE \cite{van2008visualizing}, which is a non-linear dimension reduction method for visualizing hidden vectors generated by neural networks. This method can visualize high-dimensional data in a lower-dimensional space while preserving the local structure of the original data. This process enables gathering data points with similar data distributions and separating distinct data distributions. Consequently, the confirmation of the model's proficiency in identifying disruptive phases becomes achievable by analyzing the distributions of hidden vectors computed by the models with input data, as demonstrated by \cite{zhu2020hybrid}. We conduct simulations of continuous disruption prediction in shots 30312 and 31888 with uncertainty estimation over time, which showcase the impressive results of alerting disruptions at least 500 ms before the onset of thermal quench. To elucidate the advantages of the proposed model, the investigation of uncertainty distribution among KSTAR experimental shots is conducted, leveraging the Bayesian framework to optimize the predictor's thresholds and enhance the model accuracy. 

\subsection{Disruption prediction performance}

\begin{figure}[!t]
    \centerline{\includegraphics[width=16cm, height = 10cm]{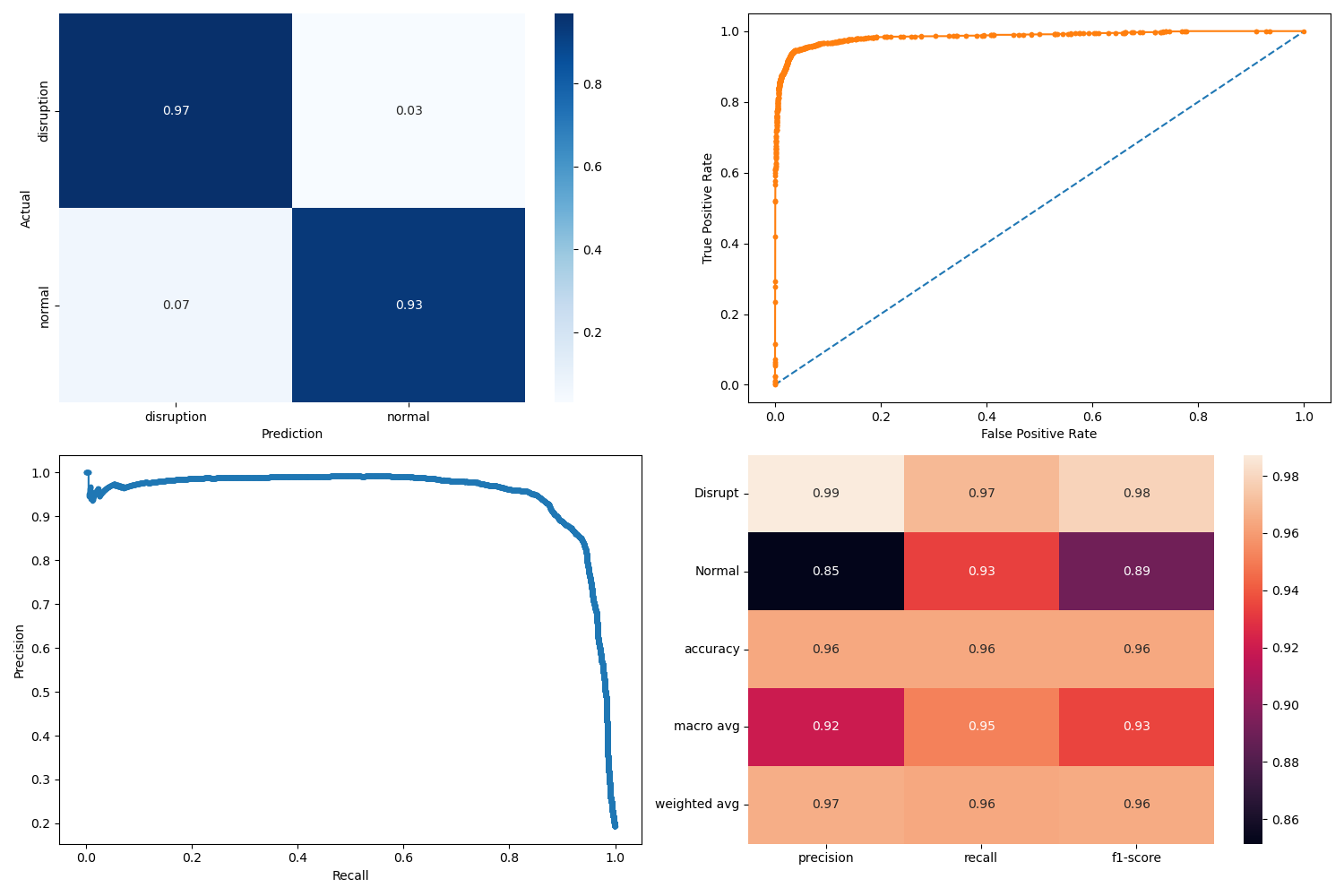}}
    \caption{Evaluation result of test data batches including confusion matrix and ROC curve}
    \label{fig:evaluation-testset}
\end{figure}

\tab To assess the model capabilities, we utilized evaluation metrics such as True positive rate (TPR), False positive rate (FPR), Precision, Recall, and F1 score with the test dataset. Given that the warning time for disruptions varies among shots, the ratio between the numbers of non-disruptive and disruptive data batches differs across experiments. This indicates that the ratio can affect the evaluation performance when assessed through test data batches. Thus, we additionally conducted a shot-by-shot evaluation by investigating simulation results of continuous disruption prediction for experimental shots. As our proposed model was trained to predict disruptions at least 40 ms before the thermal quench, a true alarm was defined when the model predicted disruptions before 40 ms from the thermal quench. Missing alarm cases include failure to forecast disruptions or delayed alarms predicted after thermal quench. These cases were treated as false alarms if alarms were alerted before the disruption warning time in experiments. The results are presented in Table \cref{table:evaluation-testset,table:evaluation-testshot,table:evaluation-testshot-metric} and Figure \ref{fig:evaluation-testset}. Table \ref{table:evaluation-testset} and Figure \ref{fig:evaluation-testset} show the evaluation result using test data batches, while Table \ref{table:evaluation-testshot} and Table \ref{table:evaluation-testshot-metric} describe the same metrics by evaluating shot-by-shot. From Table \ref{table:evaluation-testset} and Figure \ref{fig:evaluation-testset}, the macro average metrics are all over 90$\%$, and Table \ref{table:evaluation-testshot-metric} reveals that TPR is approximately 83.6$\%$ and other metrics, except FPR, also surpass 80$\%$, representing significant results in comparison to previous studies \cite{zhu2020hybrid, guo2021disruption, rea2018disruption}.

\vfill
\begin{center}
\captionof{table}{Evaluation result for test data batches}
\begin{tabular}{ |P{3cm}|P{3cm}|P{3cm}|P{3cm}| }
 \hline
    &Precision&Recall&F1 score\\
 \hline
 Disruption & 0.99 & 0.97 & 0.98\\
 Normal & 0.85 & 0.93 & 0.89\\
 \hline
 Macro average & 0.92 & 0.95 & 0.93\\
 Weighted average & 0.97 & 0.96 & 0.96\\
 \hline
\end{tabular}
\label{table:evaluation-testset}
\end{center}

\begin{center}
\captionof{table}{Evaluation result of true alarm, false alarm, and missing alarm cases}
\begin{tabular}{|P{5cm}|P{3cm}|P{3cm}|P{3cm}|}
 \hline
 \makecell{Total case\\(True alarm + Missing alarm)} & True alarm & \makecell{False alarm\\(+Early alarm)} & Missing alarm\\ [1ex]
 \hline
 61 & 53 & 9 & 8 \\
 \hline
\end{tabular}
\label{table:evaluation-testshot}
\end{center}

\begin{center}
\captionof{table}{Evaluation metrics for test shot}
\begin{tabular}{|P{1.5cm}|P{1.5cm}|P{1.5cm}|P{1.5cm}|P{1.5cm}|P{1.5cm}|}
 \hline
 TPR & FPR & Precision & Recall & F1 score & Accuracy\\
 \hline
 0.869 & 0.148 & 0.855 & 0.869 & 0.862 & 0.861 \\
 \hline
\end{tabular}
\label{table:evaluation-testshot-metric} 
\end{center}

\begin{figure}[h]
    \centerline{\includegraphics[width=12cm, height = 8cm]{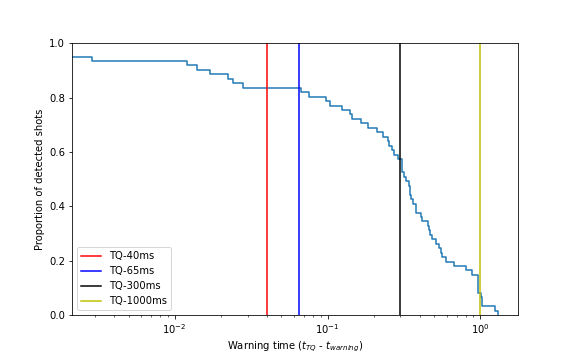}}
    \caption{Proportion of detected shot over prediction time}
    \label{fig:evaluation-cumulative}
\end{figure}

\tab In addition, the cumulative ratio of detected disruptions over the prediction time was conducted, as described in Figure \ref{fig:evaluation-cumulative}. The prediction time is the temporal difference between thermal quench and alert time. The graph reveals that most disruptions can be predicted as early as 65 ms before the thermal quench, which is earlier than the minimum required prediction time of 40 ms \cite{hollmann2015status}. However, the cumulative ratio significantly decreases when the warning time is over 300 ms. Considering the minimum required prediction time of 40 ms for mitigating disruptions in future reactors \cite{hollmann2015status}, this result suggests that our proposed model can provide disruption alerts for most cases, at least up to 40 ms preceding the onset of the thermal quench in KSTAR. 

\begin{figure}[!h]
\centering
  \begin{minipage}[b]{15cm}
    \centerline{\includegraphics[width=\columnwidth, height = 5.5cm]{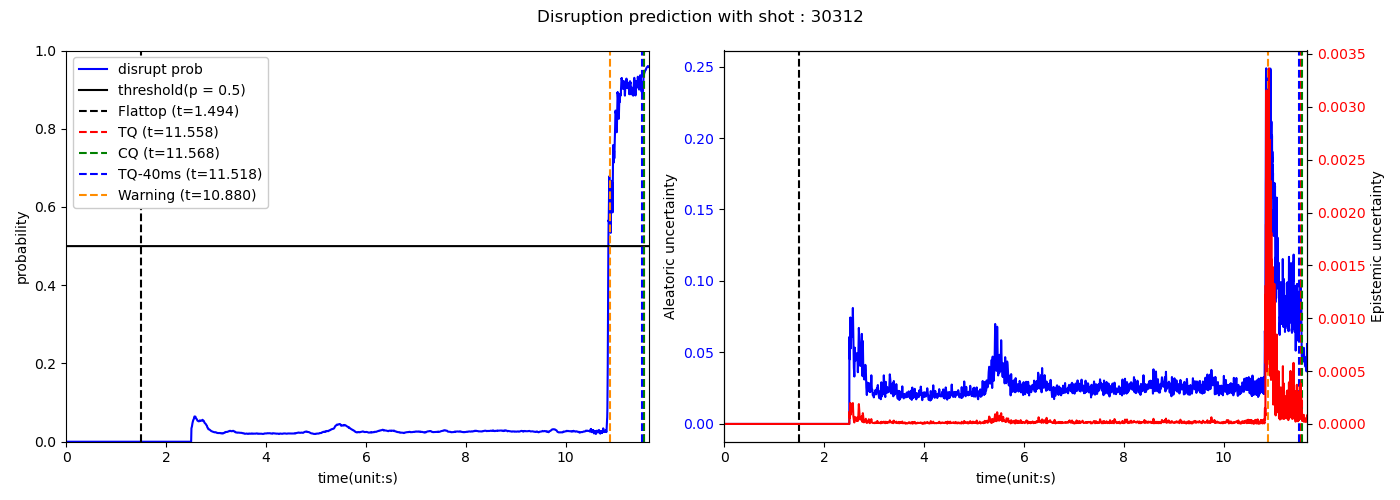}}
  \end{minipage}
  \hfill
  \begin{minipage}[b]{15cm}
    \centerline{\includegraphics[width=\columnwidth, height = 5.5cm]{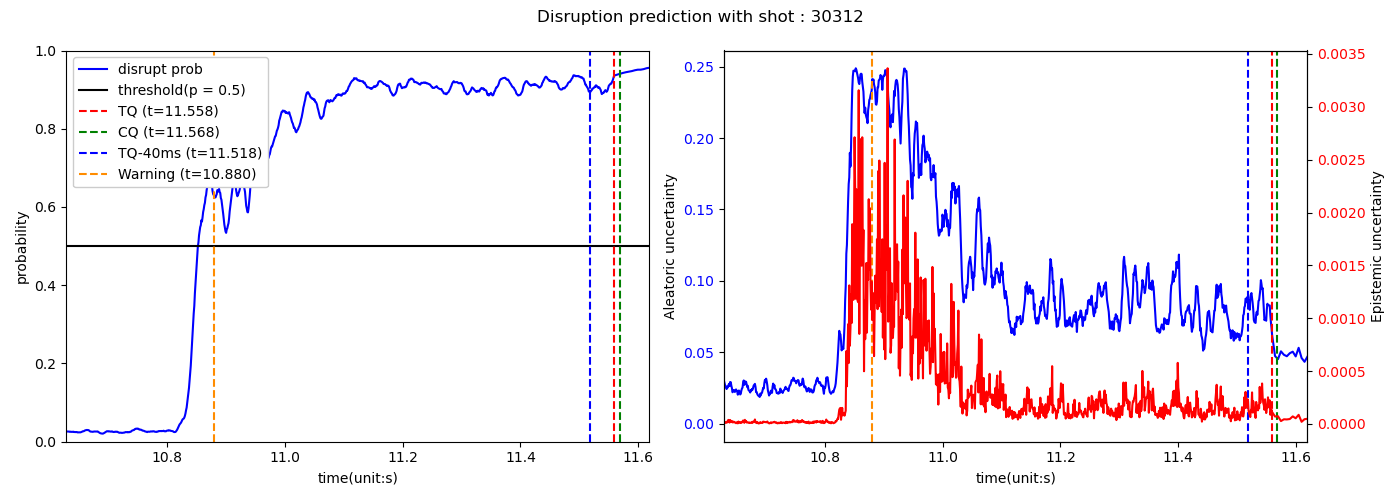}}
  \end{minipage}
\caption{The simulation results for disruption prediction 30312 are illustrated. The upper figure presents the results on the operational time scale, while the lower figure provides a zoomed-in view of the same results.}
\label{fig:continuous-disruption-prediction-shot-30312}    
\end{figure}

\begin{figure}[!h]
\centering
  \begin{minipage}[b]{15cm}
    \centerline{\includegraphics[width=\columnwidth, height = 5.5cm]{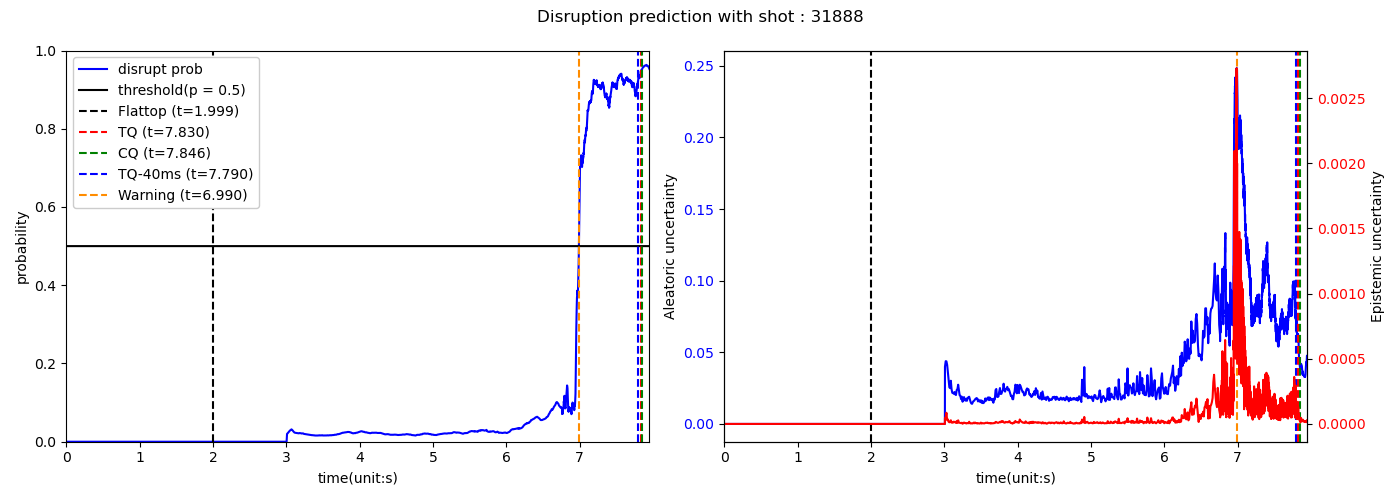}}
  \end{minipage}
  \hfill
  \begin{minipage}[b]{15cm}
    \centerline{\includegraphics[width=\columnwidth, height = 5.5cm]{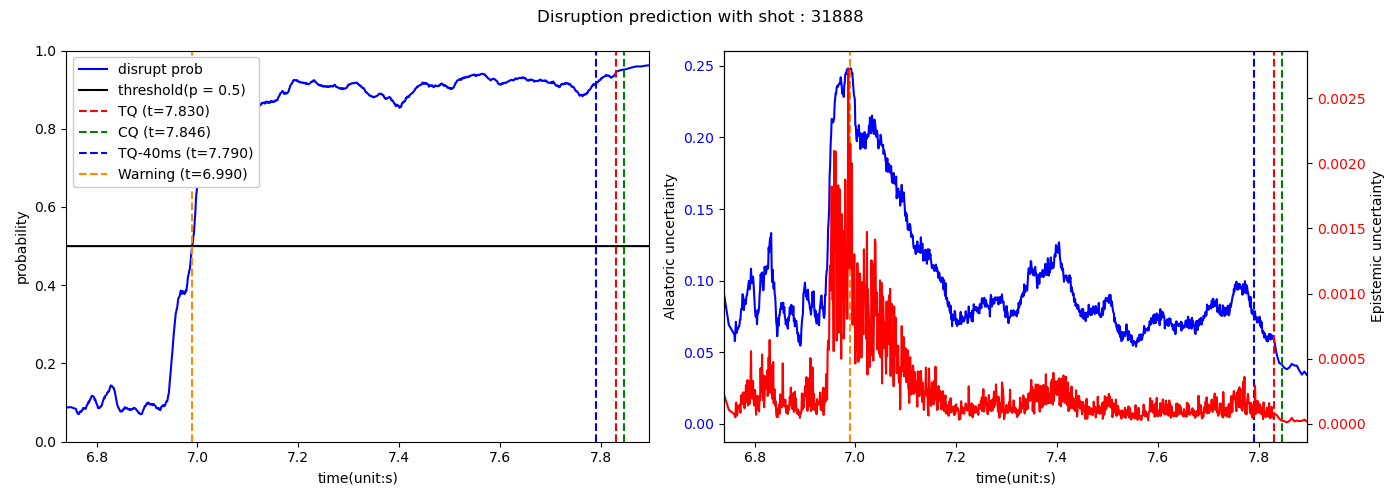}}
  \end{minipage}
\caption{The simulation results for disruption prediction 31888 are illustrated. The upper figure presents the results on the operational time scale, while the lower figure provides a zoomed-in view of the same results.}
\label{fig:continuous-disruption-prediction-shot-31888} 
\end{figure}

\tab To demonstrate the model's efficacy as a disruption warning system for tokamak devices, simulations for continuous disruption prediction were conducted with KSTAR experimental data. Specifically, shots 30312 and 31888 were selected as examples to validate our proposed model. Figure \ref{fig:shot-info-30312} and Figure \ref{fig:shot-info-31888} in Appendix B provide information on the KSTAR experiment shots 30312 and 31888, respectively. Figure \ref{fig:continuous-disruption-prediction-shot-30312} and Figure \ref{fig:continuous-disruption-prediction-shot-31888} present the simulation results of continuous disruption prediction with shots 30312 and 31888, including the uncertainty estimation over time. The model successfully predicted disruptions well before the thermal quench, occurring at approximately 680 ms and 840 ms, respectively. Notably, each probability curve exhibits no peak values before the warning time, indicating an absence of false alarms. Meanwhile, aleatoric and epistemic uncertainties drastically increase while approaching the warning regime, followed by a subsequent decrease near the thermal quench. This implies that recognizing the transition from a normal state to an unstable state near the boundary of the warning regime is initially challenging but gradually becomes evident over time. This change is particularly notable as the plasma state evolves into the thermal quench state, ultimately manifesting a distinct state with evident characteristics indicative of the disruptive phase. 

\begin{figure}[!h]
    \centerline{\includegraphics[width=8cm, height = 5cm]{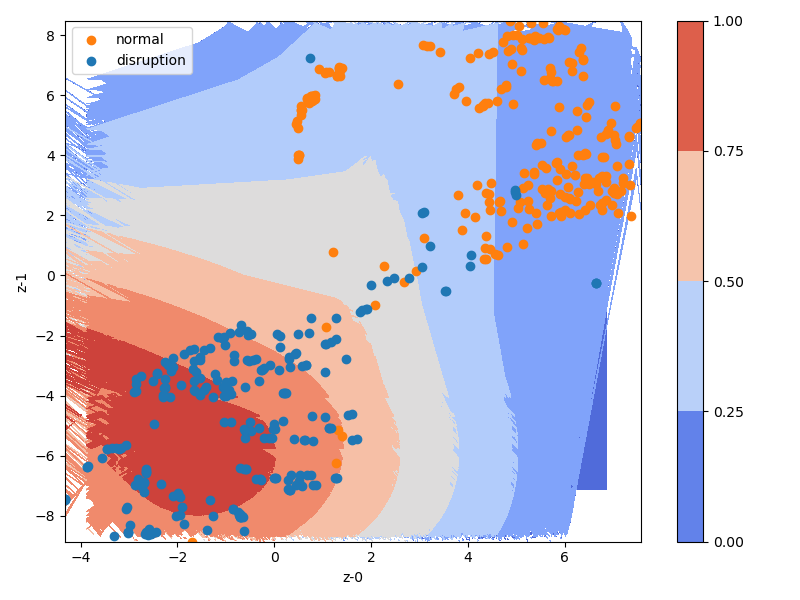}}
    \caption{Visualization of hidden vectors with test dataset by t-SNE}
    \label{fig:visualization-tSNE]}
\end{figure}

\tab To effectively analyze the model's performance across various KSTAR experimental shots, t-SNE was utilized to visualize the hidden vectors produced by the model with the test dataset. In more detail, we selectively gathered disruptive and non-disruptive data, enabling the visualization of the distribution of both disruptive and non-disruptive data. This approach offers insight for evaluating the model's capacity to identify the plasma states. A model proficient in classifying disruptive data would typically exhibit well-separated distribution between disruptive and non-disruptive KSTAR experimental data when represented through reduced hidden vectors obtained from the model. Our proposed model has demonstrated precisely this well-classified data distribution with t-SNE. Figure \ref{fig:visualization-tSNE]} shows the visualization of hidden vectors from the models using t-SNE and its 2D decision boundary. In this figure, the distribution of disruptive and non-disruptive data is clearly well-separated in 2D latent spaces, although some data points are situated in incorrect clusters, indicating missing alarm and false alarm cases. These misclassified data points should be addressed by refining the model through well-fitted training on our dataset or alternative methods, such as tuning output thresholds to compensate for the wrong decisions. We will cover this threshold-tuning method in the next section. 

\subsection{The advantages of the Bayesian neural network for predicting disruptions}
\begin{figure}[!h]
  \centering
  \begin{minipage}[b]{0.25\columnwidth}
    \centerline{\includegraphics[width=6cm, height = 5cm]{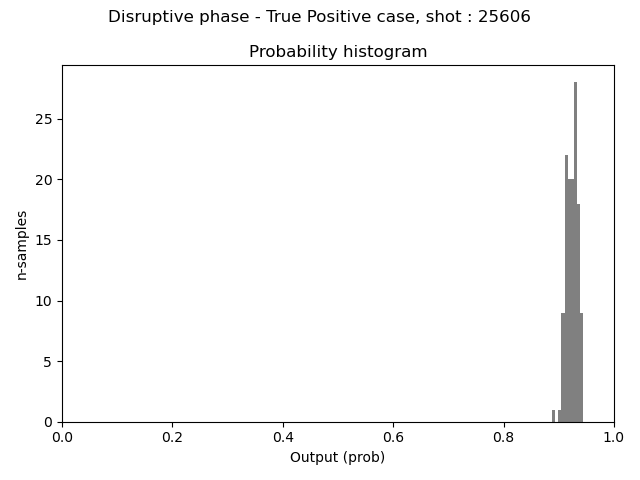}}
  \end{minipage}
  \hfill
  \begin{minipage}[b]{0.25\columnwidth}
    \centerline{\includegraphics[width=6cm, height = 5cm]{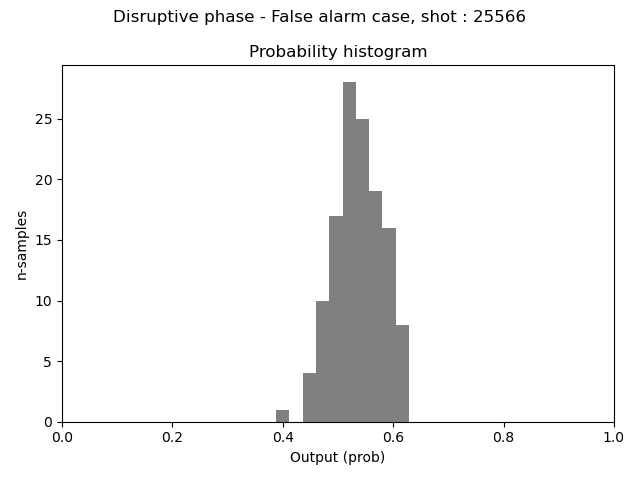}}
  \end{minipage}
  \hfill
  \begin{minipage}[b]{0.25\columnwidth}
    \centerline{\includegraphics[width=6cm, height = 5cm]{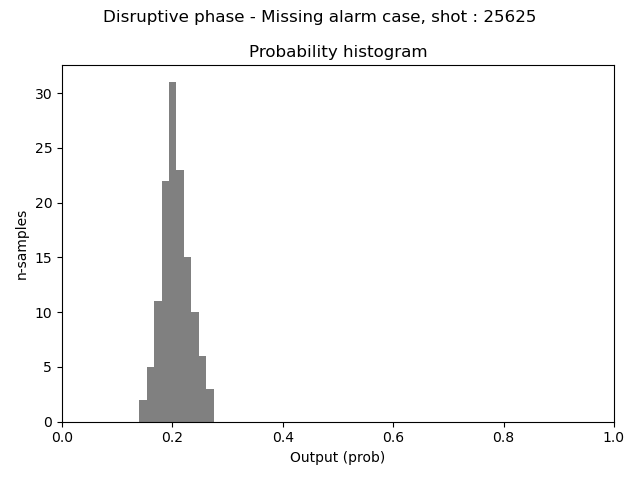}}
  \end{minipage}
  \caption{The histogram of predicted disruption probability with an example of input data for a true alarm case (left), a false alarm case (middle), and a missing alarm case (right).}
  \label{fig:uncertainty-distribution-cases}
\end{figure}

\begin{figure}[!h]
  \centering
  \begin{minipage}[b]{0.485\columnwidth}
    \centerline{\includegraphics[width=9cm, height = 6cm]{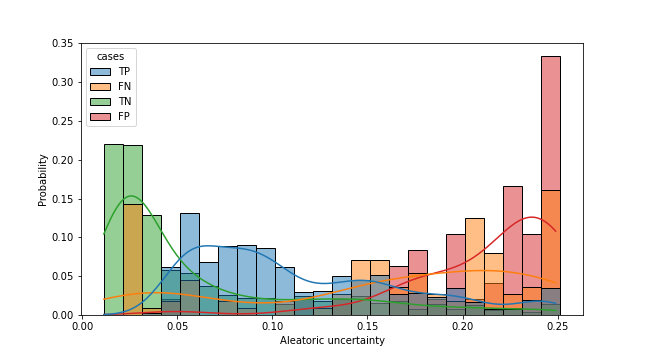}}
  \end{minipage}
  \hfill
  \begin{minipage}[b]{0.485\columnwidth}
    \centerline{\includegraphics[width=9cm, height = 6cm]{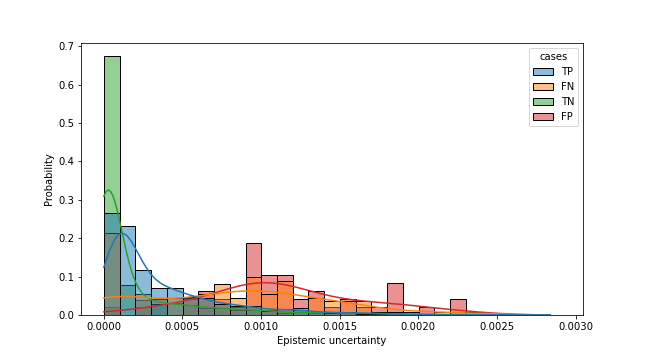}}
  \end{minipage}
  \caption{The distribution of aleatoric uncertainty (left) and epistemic uncertainty (right) of the model's predictions}
  \label{fig:uncertainty-distribution}
\end{figure}

\tab To mitigate incorrect alarm cases by utilizing uncertainty, it is crucial to analyze the uncertainty distribution for all prediction cases to capture the characteristics of each case. Figure \ref{fig:uncertainty-distribution} illustrates the aleatoric and epistemic uncertainty distribution of predictions using KSTAR experimental data. In this representation, most false alarm cases (FP) and missing alarm cases (FN) exhibit high aleatoric and epistemic uncertainty, while true alarm cases do not. This observation is corroborated by Figure \ref{fig:uncertainty-distribution-cases}, which describes the histogram of predicted disruption probability with an example of input data for a true alarm case, a false alarm case, and a missing alarm case. Table \ref{table:uncertainty-distribution-cases} represents the estimation of aleatoric and epistemic uncertainty corresponding to Figure \ref{fig:uncertainty-distribution-cases}. The input data were generated as repeated tensors along the batch dimension, with a size equal to the sample size set at 128, and the output probability was computed. The stochastic nature of the model weights induces variability, resulting in a distribution with discernible deviation. In Figure \ref{fig:uncertainty-distribution-cases}, true alarm cases have a low deviation, while false and missing alarm cases have a high deviation. As the deviation of the output probability distribution represents the predictive uncertainty, expressed as the summation of aleatoric and epistemic uncertainty, this result aligns with Figure \ref{fig:uncertainty-distribution}. This indicates that both data and model uncertainty contribute to the failure to predict disruptions, as evident in the cases of incorrect alarms with high uncertainty. Nevertheless, insights obtained from the analysis of uncertainty distribution results suggest the feasibility of reducing incorrect alarms by fine-tuning the thresholds for uncertainty. 

\begin{center}
\captionof{table}{The aleatoric and epistemic uncertainty of predicted output with an example of input data for a true alarm case, a false alarm case, and a missing alarm case.}
\begin{tabular}{|P{3.5cm}|P{2cm}|P{2cm}|P{2cm}|}
 \hline
    &True alarm & False alarm& Missing alarm \\
 \hline
 Aleatoric uncertainty & 0.0676 & 0.2474 & 0.1668 \\
 Epistemic uncertainty & 0.0001 & 0.0016 & 0.0008 \\
 \hline
\end{tabular}
\label{table:uncertainty-distribution-cases}
\end{center}

\tab Typically, the model predicts disruptions when the output probability surpasses the threshold, which is conventionally set to 0.5. However, fine-tuning the model's sensitivity for detecting disruptions is achievable by adjusting the threshold of the output probability. An increase in the threshold results in a reduction in the false alarm rate and an increase in the missing alarm rate. Conversely, a decrease in the threshold increases the false alarm rate with a simultaneous decrease in the missing alarm rate. This implies that tuning the threshold of output probability alone entails a trade-off between precision and recall. The condition for determining disruptive data with a given output probability and threshold is expressed as Equation \ref{eq:criteria-original}. 

\begin{equation}
\hat{y}=\begin{cases}
1 & p(y|x) \ge p_{thres} \\
0 & \mbox{otherwise}
\end{cases}
\label{eq:criteria-original}
\end{equation}

where $p(y|x)$ is the output probability computed by the model and $p_{thres}$ is the threshold for the output probability. Labels 0 and 1 correspond to non-disruptive and disruptive, respectively. In this study, we employed threshold tuning for the output probability and the uncertainties. Note that the false alarm rate can be reduced without a large increase in the missing alarm rate through additional tuning of the uncertainty threshold. Equation \ref{eq:criteria-new} formulates the criteria for determining whether the data is disruptive. 

\begin{equation}
\hat{y}=\begin{cases}
1 & p(y|x) \ge p_{thres} \mbox{ and } \sigma \le \sigma_{threshold} \\
0 & \mbox{otherwise}
\end{cases}
\label{eq:criteria-new}
\end{equation}

\begin{figure}[h]
  \centering
  \begin{minipage}[b]{0.485\columnwidth}
    \centerline{\includegraphics[width=10cm, height = 7cm]{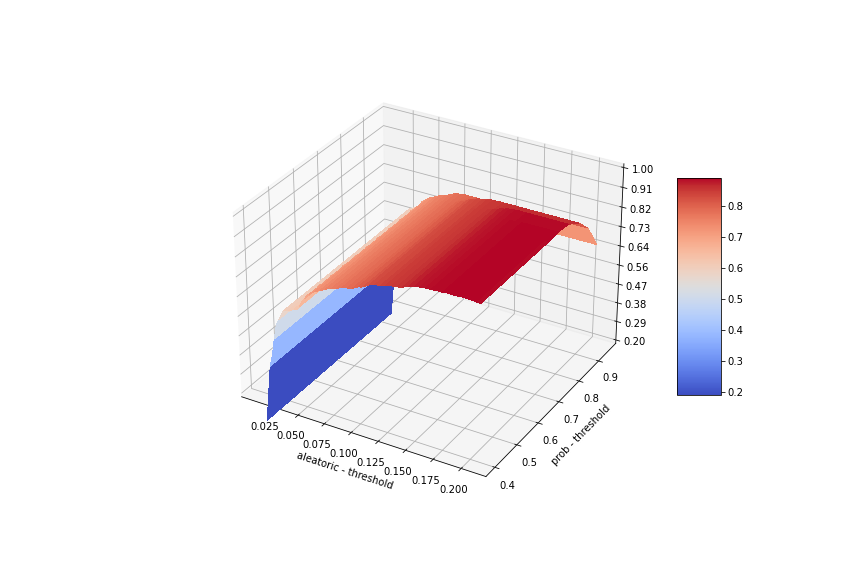}}
  \end{minipage}
  \hfill
  \begin{minipage}[b]{0.485\columnwidth}
    \centerline{\includegraphics[width=10cm, height = 7cm]{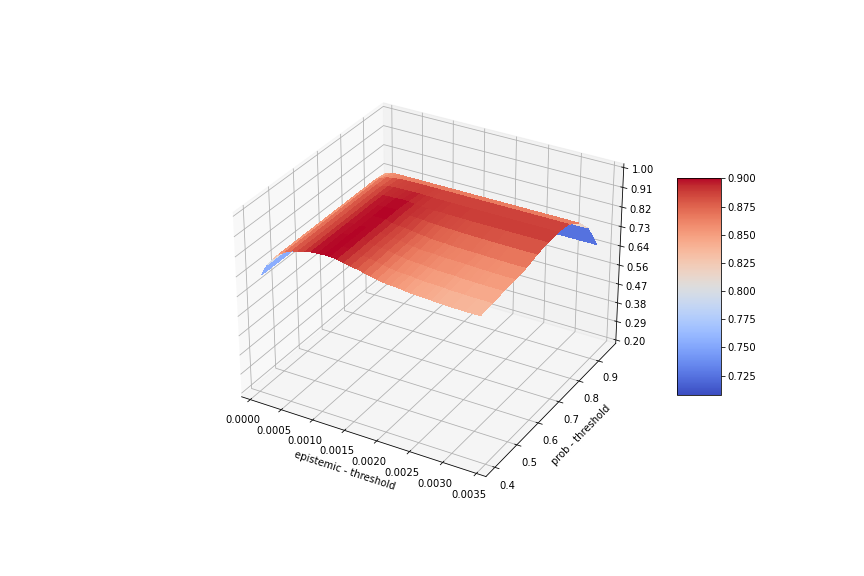}}
  \end{minipage}
  \caption{Exploring 3D parametric space for optimizing model accuracy: This figure illustrates F1 score optimization through threshold tuning. The left figure explores tuning output probability and aleatoric uncertainty threshold, while the right figure focuses on tuning output probability and epistemic uncertainty.}
  \label{fig:uncertainty-optimization}
\end{figure}

$\sigma$ and $\sigma_{threshold}$ indicate the uncertainty and its threshold, respectively. We performed three optimization processes to enhance the model's accuracy. The initial case involved tuning the threshold for output probability. In the second case, we tuned the threshold for both output probability and aleatoric uncertainty. The last case focused on tuning the threshold for output probability and epistemic uncertainty. The results are shown in Table \ref{table:uncertainty-optimization} and Figure \ref{fig:uncertainty-optimization}. Table \ref{table:uncertainty-optimization} compares metrics between the original and optimized cases. Figure \ref{fig:uncertainty-optimization} describes exploring 3D parametric space to find the optimal threshold configuration for the model's accuracy. Interestingly, in the last case, the threshold tuning for output probability and epistemic uncertainty shows a significant improvement in precision with a smaller recall reduction compared to the other cases. Typically, tuning the output probability threshold induces a trade-off between precision and recall. However, incorporating the uncertainty threshold enables enhancing the model precision without experiencing significant degradation in recall, thereby mitigating the inherent trade-off relationship. In essence, the capability to quantify the uncertainty within a Bayesian framework, facilitating reliable predictions by incorporating uncertainty, becomes imperative for forecasting disruptions while mitigating the occurrence of false alarms.

\begin{center}
\captionof{table}{The result for optimizing model accuracy through threshold tuning.}
\begin{tabular}{|P{4cm}|P{2cm}|P{2cm}|P{2cm}|P{2cm}|P{2cm}|}
 \hline
    &Threshold (Probability)&Threshold (Uncertainty)&Precision&Recall&F1 score\\
 \hline
 w/o tuning & 0.500 & - & 0.819 & 0.878 & 0.848 \\
 \hline
 Prob only & 0.750 & - & 0.960 & 0.830 & 0.890 \\
 Prob + Aleatoric & 0.550 & 0.193 & 0.956 & 0.834 & 0.891 \\
 Prob + Epistemic & 0.650 & 0.001 & 0.958 & 0.853 & 0.902 \\
 \hline
\end{tabular}
\label{table:uncertainty-optimization}
\end{center}

\begin{figure}[!h]
\centering
  \begin{minipage}[b]{15cm}
    \centerline{\includegraphics[width=\columnwidth, height = 5.5cm]{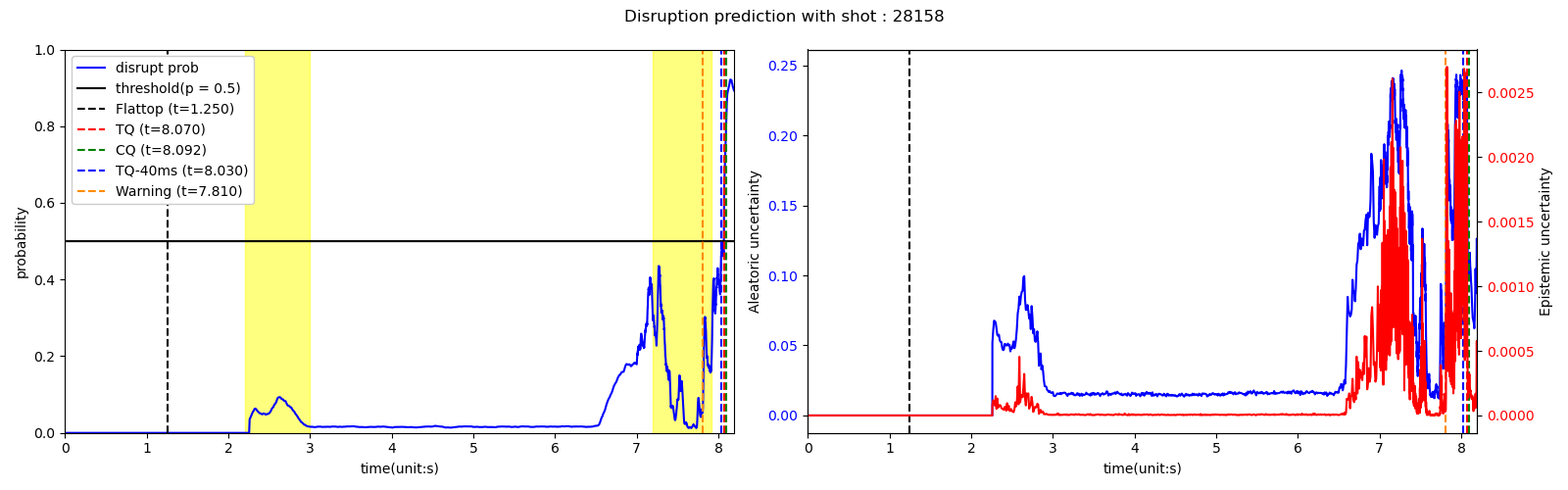}}
  \end{minipage}
  \hfill
  \begin{minipage}[b]{15cm}
    \centerline{\includegraphics[width=\columnwidth, height = 5.5cm]{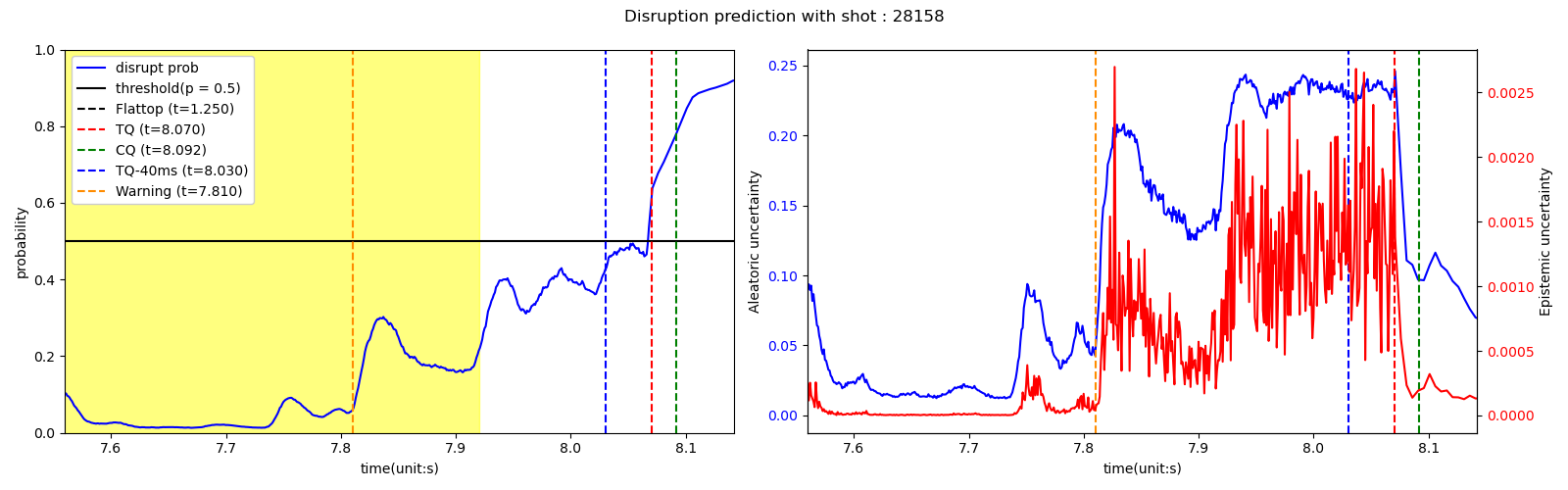}}
  \end{minipage}
\caption{The simulation results for disruption prediction 28158 are illustrated. The upper figure presents the results on the operational time scale, while the lower figure provides a zoomed-in view of the same results. The yellow background represents the region where false alarms occur.}
\label{fig:continuous-disruption-prediction-shot-28158}    
\end{figure}

\tab This improvement was further evidenced by the simulation conducted for continuous disruption prediction with KSTAR experimental shot 28158. The details of the shot are illustrated in Figure \ref{fig:shot-info-28158} in Appendix B. False alarms in this shot are effectively filtered by utilizing threshold tuning for both the output probability and epistemic uncertainty, as described in Figure \ref{fig:continuous-disruption-prediction-shot-28158}. This simulation result suggests the potential for enhancing a reliable disruption alarm system through uncertainty estimation. 

\section{Conclusion}

\tab In this paper, we developed an advanced model based on a Bayesian neural network capable of predicting disruptions and estimating the uncertainty of its prediction. By leveraging 0D plasma parameters from EFIT and diagnostic data, our model excels in covering multi-time scale input signals through TCN architecture, showcasing the advanced warnings with a prediction time over 40 ms preceding the onset of thermal quench. The capability to compute aleatoric and epistemic uncertainty with predictions suggested the possibility of achieving reliable disruption forecasts by optimizing thresholds for the output probability and uncertainty. Our comparison analysis of uncertainty distribution revealed that missing and false alarms exhibit high uncertainty compared to true alarm cases. This underscored the crucial role of tuning thresholds on uncertainties in enhancing the prediction accuracy. The practical application was demonstrated by implementing the post-optimization process of tuning thresholds with test data. 

\tab In the future, the following topics should be focused on. First, a real-time disruption prediction should be conducted on various tokamak devices, including KSTAR, to validate the model's capability of predicting disruptions. This requires model acceleration techniques \cite{cheng2018model}. Second, the physical consistency should be desired. In other words, physics-combined methods are necessary to enhance interpretability. A physics-informed neural network \cite{raissi2019physics} is one of the solutions for improving interpretability, which is also applied to plasma research \cite{van2020fast,mathews2021uncovering,abramovic2022data,rossi2023potential}. Finally, online learning \cite{hoi2021online} should be adapted for future research. Given that real-time adaptation needs the model to follow the new data pattern induced by the time-evolving dynamic environment, it is necessary to update the disruption predictor as new data arrives. An online learning algorithm allows the model to capture these dynamic patterns. These challenges will be covered in future work. 

\bibliographystyle{unsrt}
\bibliography{references}

\begin{appendices}
\section{The architecture of our proposed model}
\begin{figure}[h]
    \centerline{\includegraphics[width=14cm, height = 14cm]{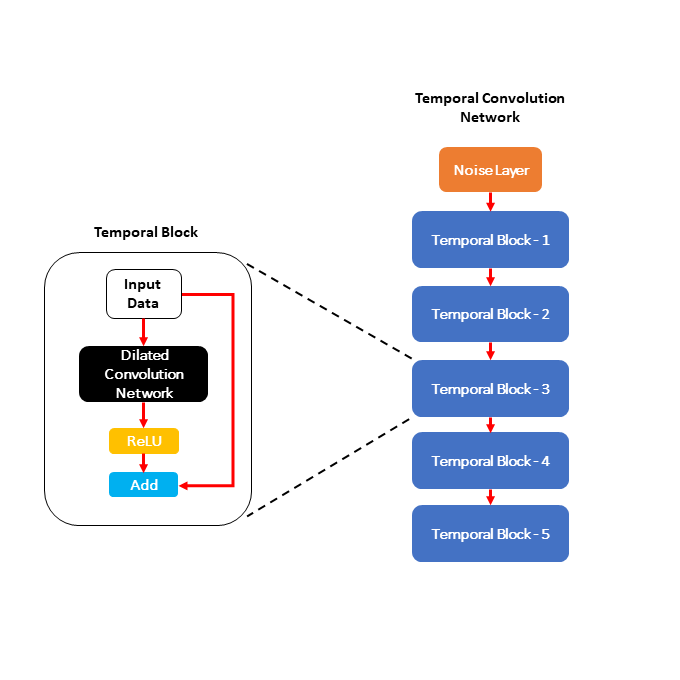}}
    \caption{The architecture of Temporal Convolution Block and Temporal Convolution Network (TCN)}
    \label{fig:TCN-architecture}
\end{figure}

\begin{figure}[h]
    \centerline{\includegraphics[width=18cm, height = 18cm]{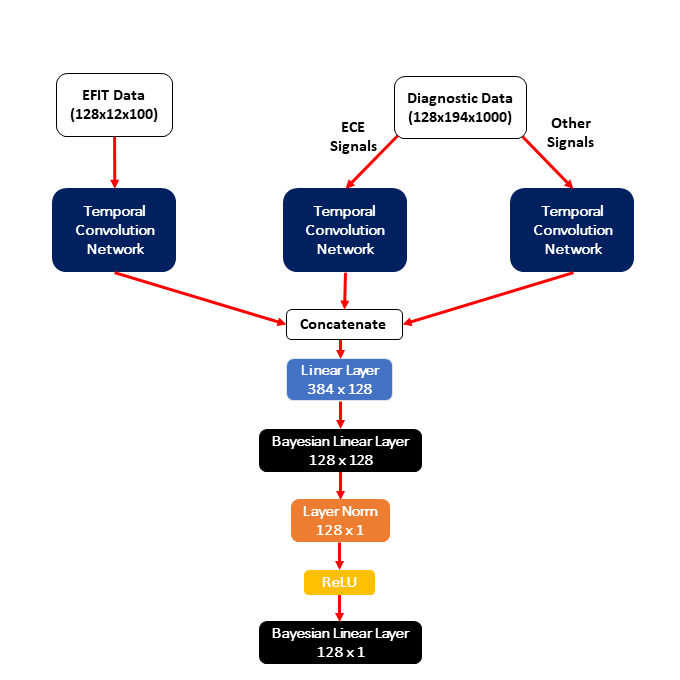}}
    \caption{Our proposed model's architecture}
    \label{fig:model-architecture}
\end{figure}

\tab This section describes the detailed architecture of our proposed model containing TCN and Bayesian neural networks. Figure \ref{fig:TCN-architecture} represents the simplified structure of TCN containing five sequential Temporal Blocks. The residual connection and dilated convolution layer are included in each Temporal Block. Figure \ref{fig:model-architecture} describes the overall architecture of our proposed model. There are three TCN modules as independent temporal encoders. We split ECE signals from diagnostic data since ECE signals need additional data refinement to avoid numerical instability during the training process, thereby adding another TCN module to process ECE signals. 


\section{Information of KSTAR experimental shots 28158, 30312, and 31888}
\begin{figure}[h]
    \centerline{\includegraphics[width=\columnwidth, height = 7cm]{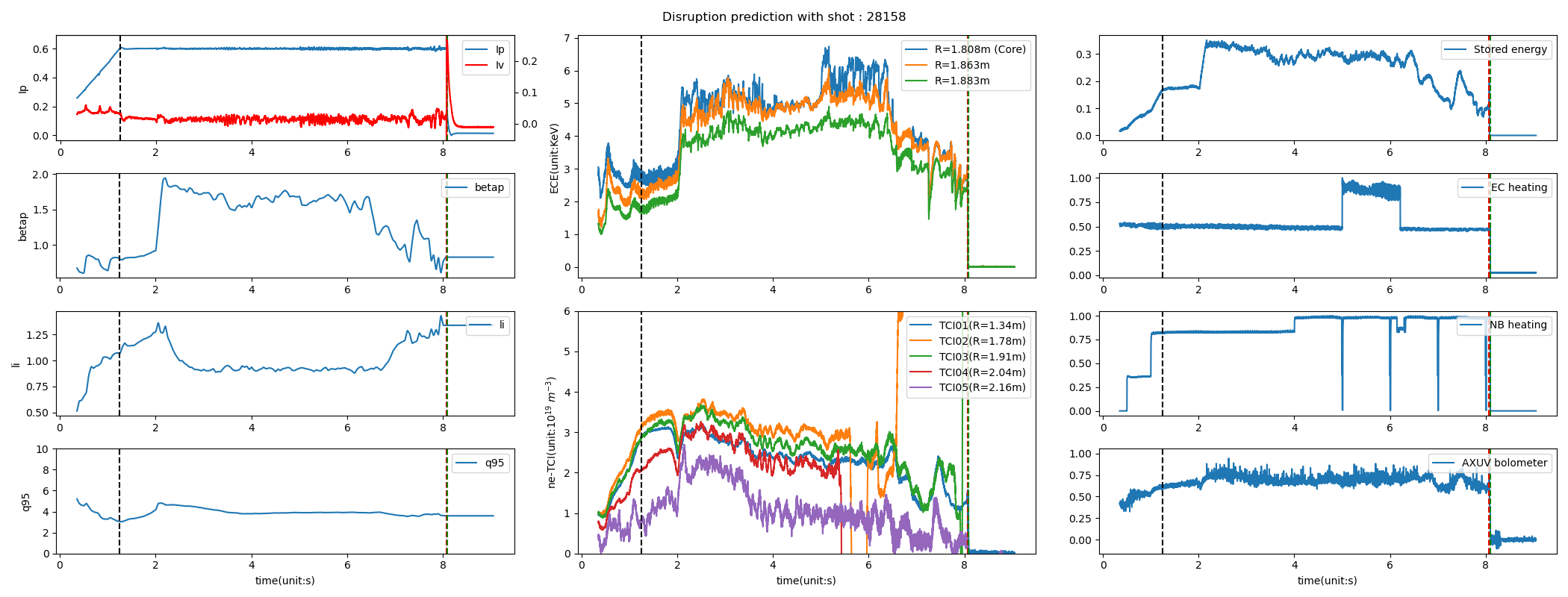}}
    \caption{KSTAR experimental shot 28158}
    \label{fig:shot-info-28158}
\end{figure}

\begin{figure}[h]
    \centerline{\includegraphics[width=\columnwidth, height = 7cm]{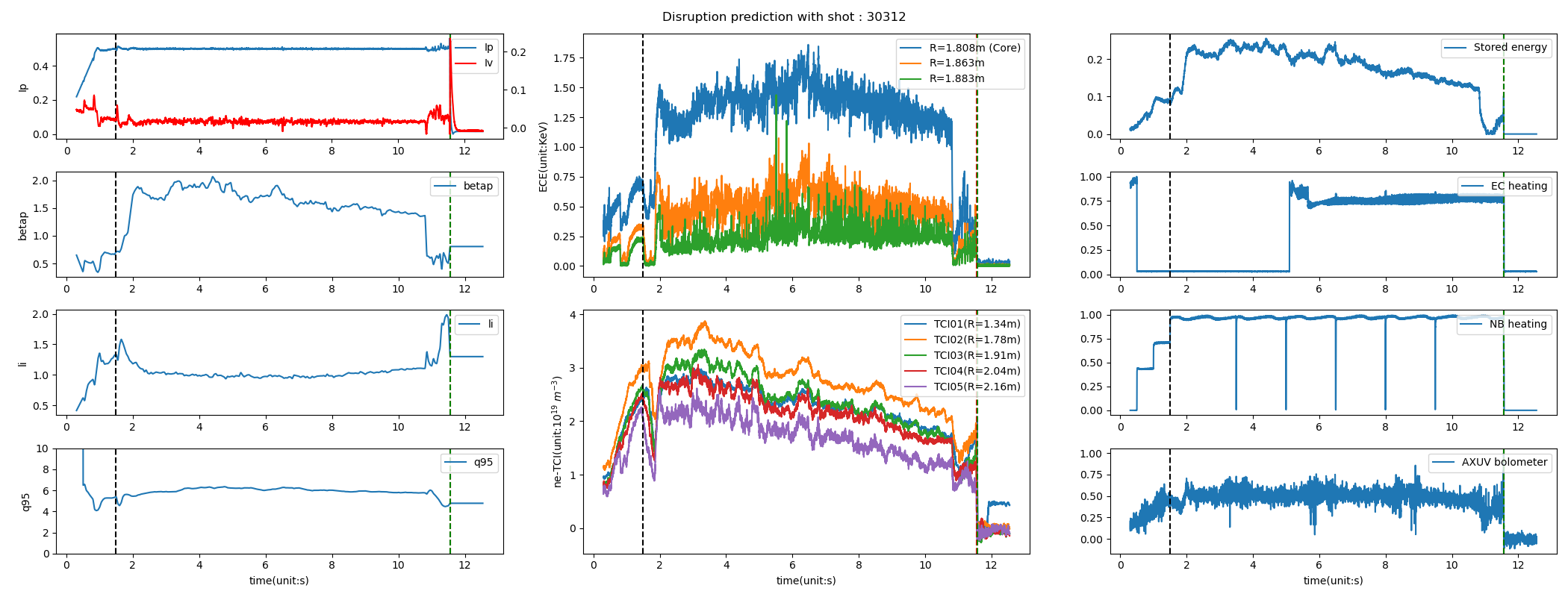}}
    \caption{KSTAR experimental shot 30312}
    \label{fig:shot-info-30312}
\end{figure}

\begin{figure}[h]
    \centerline{\includegraphics[width=\columnwidth, height = 7cm]{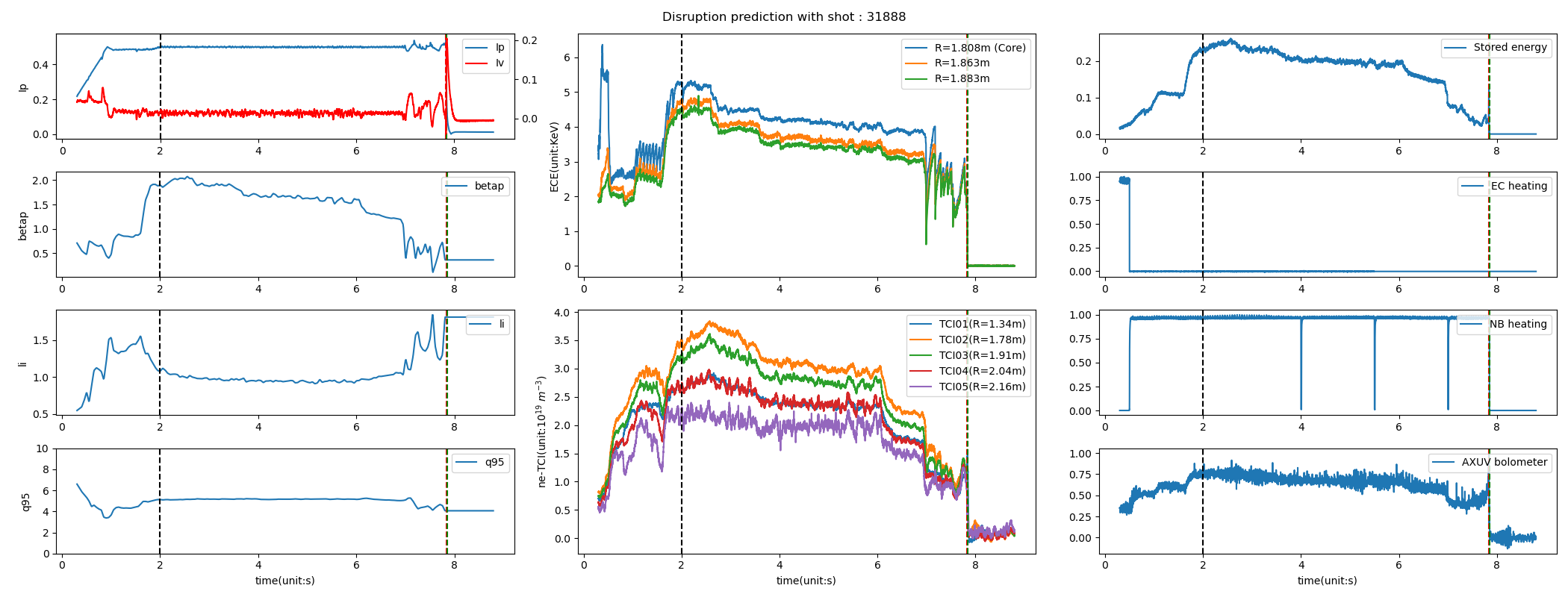}}
    \caption{KSTAR experimental shot 31888}
    \label{fig:shot-info-31888}
\end{figure}

\tab This section describes the details of KSTAR experimental shots 28158, 30312, and 31888. Figure \ref{fig:shot-info-28158, fig:shot-info-30312, fig:shot-info-31888} represent EFIT parameters, including $\beta_n$, internal inductance and $q_{95}$, and diagnostics such as plasma current, vessel current, ECE, TCI, plasma stored energy, EC heating power, NB heating power, and UV photodiode bolometer.

\end{appendices}
\end{document}